\documentclass[twocolumn,aps,prl,showpacs,preprintnumbers,amsmath,amssymb]{revtex4}
\usepackage{setspace}
\begin{document}

\title {On the Topological Origin of Entanglement in Ising Spin Glasses}

\author{V. V. Sreedhar\footnote{\sl sreedhar@iitk.ac.in}}

\affiliation{Department of Physics\\
Indian Institute of Technology\\
Kanpur, 208016 India}

\begin{abstract}
The origin of thermal and quantum entanglement in a class of three-dimensional 
spin models, at low momenta, is traced to  purely topological reasons. The 
establishment of the result is facilitated by the gauge principle which, when 
used in conjunction with the duality mapping of the spin models, enables us to 
recast them as lattice Chern-Simons gauge theories. The thermal and quantum 
entanglement measures are expressed in terms of the expectation values of 
Wilson lines, loops, and their generalisations. For continuous spins, these are
known to yield the topological invariants of knots and links. For Ising-like 
models, they are expressible in terms of the topological invariants of 
three-manifolds obtained from finite group cohomology -- the so-called 
Dijkgraaf-Witten invariants. 
\end{abstract}


\maketitle

The history of modern physics is replete with stories of the progress that
followed every time the notion of instantaneous action at a distance between
two objects was sacrificed in favour of the concept of an interaction between 
the participants. With an appropriate choice of gauge fields acting as 
mediators of the interaction, we can account for most of what is  known about 
the fundamental forces of nature. Several examples readily come to mind 
{\it viz.} Maxwell's theory of electromagnetism, the standard model of the 
electroweak forces, quantum chromodynamics -- the theory of strong, nuclear 
forces -- and Einstein's general theory of relativity for gravitational forces 
\cite{lor}. Drawing inspiration from these examples, we shall use the idea of 
gauge invariance as a {\it tour de force} to gain insight into the origin of 
entanglement in Ising spin systems.  

The need to understand the origin of quantum entanglement arises from the 
recognition that it is the theoretical bedrock that supports the emerging 
revolution in storing, processing, and retrieving information \cite{chuang}.
Most of the work on the physics of quantum entanglement is -- in view of its 
aforementioned technological import --  focussed on finding ways to quantify 
it, in developing sophisticated algorithms to handle computational complexity, 
in grappling with the issue of decoherence, and last but not least, in 
producing experimental devices which shape the promised revolution. Under the 
circumstances, very little time is invested in finding an answer to a 
fundamental question: What is the origin of entanglement in quantum systems? 
We will show that the gauge principle once again holds the key to unlock this, 
yet another, mystery of nature. 

In view of the fact that entanglement is a purely quantum mechanical feature, 
unlike the fundamental forces of nature alluded to earlier, it may seem that
the gauge principle is inapplicable in this context. Such apprehensions are 
unfounded because it is by now well-known that purely quantum features with no 
classical analogues are often associated with non-trivial topological aspects 
of the theory. The simplest example which illustrates this intimate relation is
that of a free nonrelativistic point particle moving on a circle. A total 
derivative term added to the Lagrangian of this model leaves the classical 
mechanics of the particle unchanged, but leads to different, inequivalent 
quantizations of the particle due to the topology of the circle \cite{wilczek}. 
Another example concerns identical particles which acquire anyonic quantum 
statistics in a plane, due to the non-trivial topology of the configuration 
space \cite{leinaas}. The important role played by topological objects like 
monopoles and vortices in producing confinement, and instantons in producing 
anomalies, in Yang-Mills gauge theories, is also well documented \cite{jackiw}.
The reasons for the quantization of charge in electrodynamics, flux in 
superconductivity and conductance in the Hall effect can all be traced to 
their topological antecedents \cite{thouless}. It should be noted that 
for each of these examples, there is no classical counterpart; the physical 
effects have a topological origin, and are most elegantly captured by the gauge 
principle. We then conjecture that quantum entanglement is also purely 
topological in origin and substantiate this claim by using the gauge principle,
in the concrete example of the three-dimensional Ising model.  

We mention that an intriguing analogy between quantum entanglement and 
classical topology was first publicised by Arvind \cite{arvind}, following his 
serendipitous discovery of the similarity between the entanglement properties 
of the Greenberger-Horne-Zeilinger state \cite{ghz} and the curious linking 
properties of Borromean and Hopf rings. The analogy was further developed 
by Kauffman and Lomonaco \cite{kauff} who examined the parallels between 
entanglement of states in quantum mechanics on the one hand, and the strings of
a braid on the other. The hope that the  breakthrough in finding new 
topological invariants of knots and links \cite{witten} may be used to 
characterize/classify entangled quantum states through the above analogy, 
gave rise to some initial excitement. The limitations of such a hope, as 
delineated in \cite{arvind}, were primarily because the much-needed rule to 
associate a closed loop to a quantum state, and the snipping of a loop to a 
measurement, could not be formalised. In the present work we would like to 
approach this problem from Heisenberg's point of view, and deal directly with 
the physically observable quantities. This approach enables us 
to circumvent the need to define the topological equivalents of spins and 
measurements. We thus overcome the temptation to look for parallels between 
quantum entanglement and classical topology. Instead, we search for the 
intrinsic topological content of entanglement in a prototypical spin system 
namely, the three--dimensional Ising model. In the process, we discover 
unambiguous relations between the usual measures of entanglement \cite{neumann}
and the expectation values of gauge--invariant observables in a $Z_2$ lattice 
Chern-Simons gauge theory. As is well-known, the latter are just the 
observables used to calculate topological invariants of knots and links in 
topological field theories with continuous groups \cite{kaul}. For the case of 
finite groups, the appropriate topological invariants, obtained by using finite
group cohomology, are called the Dijkgraaf-Witten invariants \cite{dw}.

The Ising model is, as usual, defined by the Hamiltonian $H = - J
\sum_{<ij>}S_iS_j$,  where $J>0$ and $S_i =\pm 1$. Here $i,j$ label the sites 
of a three-dimensional cubical lattice and the $<>$ parantheses indicate that 
the summation is over different, but nearest neighbour sites. The positivity of
$J$ implies that the ferromagnetic state minimises the energy. For the opposite
sign of $J$, the antiferromagnetic state would be favoured. As already 
motivated, we proceed to replace the above Hamiltonian which couples spins at 
different sites, by  
\begin{equation}\label{using}
H_U =  - J \sum_{<ij>}S_iU_{ij}S_j 
\end{equation}
in which the interaction between spatially separated spins is mediated by the 
gauge field $U$. The $U_{ij}$ live on the links connecting sites $i$ and $j$, 
are $Z_2$--valued, and hence equal to $\pm 1$. The gauged Ising model presented
above accommodates either a ferromagnetic or an antiferromagnetic bond between 
various nearest neighbour sites \cite{fradkin}. Such models play a crucial role
in understanding the behaviour of disordered systems; in which subject they are
referred to, and extensively studied, as {\it spin glasses} \cite{debu}.  In 
this letter, however, a spin glass is merely an expedient to replace the action
at a distance (of the order of the lattice spacing) between Ising spins by an 
interaction mediated by the gauge fields $U_{ij}$.  

We define the partition function for the above model in the usual way as the 
trace over the Gibbs' measure {\it i.e.} $Z = {\hbox{Tr }}~e^{-\beta H_U}$
where $\beta = 1/k_BT$, $k_B$ being the Boltzmann constant, and $T$, the 
temperature. In the first step, called {\it quenching} in the language of spin 
glasses, we sum over all the spin degrees of freedom to get \cite{senthil} 
\begin{equation}\label{trace}
Z =  A(J) \sum_{U,V} e^{-S_V - S_{CS}} 
\end{equation}
where 
\begin{equation}\label{dual}
S_V = -{\tilde J}\beta \sum_{\Box}\prod_{\Box} V_{ij}
\end{equation}
and 
\begin{equation}\label{chern-simons}
S_{CS} = \beta\sum_{<ij>}i{\pi\over 4}(1 - \prod_\Box V)(1 - U) 
\end{equation}
The constants $A$ and $\tilde J$ are defined in terms of 
the strength of the magnetic interaction $J$, as follows:
\begin{equation}\label{couplings}
A(J) = {\hbox{cosh}}J \qquad {\hbox{and }}\qquad {\hbox{tanh}}{\tilde J} = 
e^{-2J} 
\end{equation}
The $V_{ij}$, like the $U_{ij}$, are $Z_2$-valued fields, but live on the 
links of the dual lattice obtained by a rigid translation of each site on 
the primary lattice by half a lattice spacing uniformly along each of the 
coordinate axes. The product of the dual gauge variables $V_{ij}$ around 
an elementary plaquette on the dual lattice is indicated by the $\Box$ 
under the product symbol. When the $\Box$ appears under the summation symbol, 
it is an instruction to sum over all such elementary plaquettes in the dual 
lattice. The $S_V$ term is readily recognised as the standard Wilsonian 
action for the gauge field $V$. The $S_{CS}$ term is physically a measure of 
the flux passing through a dual plaquette perpendicular to a given link $U$ on 
the primary lattice. It is a Chern-Simons action on the lattice. Two crucial 
steps in arriving at the above result consist in an expansion of $Z$ in the 
characters of the $Z_2$ group, and the introduction of the dual lattice 
variables. The details can be found in \cite{senthil}. The generalisation to 
other finite abelian groups $Z_p$ follows along the same lines \cite{savit}, 
as does the limiting case $p\rightarrow\infty$. The latter corresponds to the 
familiar lattice $U(1)$ Chern-Simons theory \cite{adams}.   

At this stage, a few general observations regarding the origin of the various 
terms in (2) are in order. First, from the continuum perspective, for the case 
of a continuous group, this result is easily anticipated. It is well-known from
 \cite{deser} that a derivative expansion of the fermionic determinant in this 
case, obtained by integrating out the fermions coupled to a background gauge 
field in three-dimensional spacetime, produces at the lowest two orders, the 
Chern-Simons and Maxwell terms. The result in (2-5) is a lattice realisation of
the above celebrated continuum result. Second, from a purely lattice point of 
view, it is well-known that the three-dimensional Ising model is equivalent to 
a $Z_2$ lattice gauge theory on the dual lattice \cite{wegner}. The Ising spins 
with nearest neighbour interactions on the primary lattice have thus been 
traded for the $V$-fields with a Wilson action that appears in (3). In the 
duality transformation, the $U$ fields introduced by the gauge principle on 
the primary lattice are mere spectators and hence we arrive at a set of two 
$Z_2$-valued fields. Moreover, since there is an inextricable linkage between 
the primary and dual lattices, each link on the primary lattice pierces a 
plaquette of the dual lattice (and {\it vice versa}), and the Chern-Simons 
action is precisely a measure of this flux. Third, the noticeable difference 
between the continuum and lattice realisations of what is essentially the same 
result is reminiscent of the lattice fermion doubling problem and has been 
discussed before in attempts to discretise Chern-Simons gauge theories 
 \cite{sen}. It may be recalled for the sake of completeness that, the 
mathematical reason for the doubling is that the Hodge star operator appearing 
in the Chern-Simons term couples cochains of a simplicial decomposition with 
the cochains of the dual decomposition. From a physical point of view, the 
Chern-Simons term, owing to the fact that it has no independent gauge-invariant
dynamics, couples matter fields to the magnetic flux. In the present case, the 
Ising spins residing on the sites of the primary lattice are the matter fields.
The magnetic flux, in lattice gauge theory, is defined by the gauge fields on 
plaquettes. There is no natural way of coupling these two objects without doing
violence to the structure of the lattice. This very fact was used by Kantor and
Susskind for one of the early models for anyons \cite{kantor}. 

Finally, we mention that models of topological quantum computation using anyons
were proposed in \cite{kitaev} and experiments with flux-qubits \cite{delft} 
are beginning to realise some of these ideas. Quite independently, remarkable 
progress has been achieved by using statistical mechanical techniques to 
study spin glasses in theories of information processing \cite{nishi} 
following the proposal of \cite{sourlas}. Interestingly, equation (2) 
establishes a connection between the above two seemingly different approaches 
to the subject of quantum information. More importantly, since the conceptual 
roots of the former approach lie in topology, while those of the latter lie in 
gauge invariance, it reinforces the idea anticipated at the beginning of 
this letter.   
  
To proceed further, we note that although equation (4) does not treat $U$ and 
$V$ on the same footing, the exponential of $-S_{CS}$ which appears in equation
(2) is invariant under an exchange of $U$ and $V$ \cite{senthil}. However, 
equation (2) itself is lopsided because it does not have a Wilson term 
associated with the $U$ field. This suggests that in the next step, called 
{\it configurational averaging}, where the quenched gauge degrees of freedom 
are summed over subject to a given distribution, we choose the weights such 
that the $U\leftrightarrow V$ symmetry is restored. This is tantamount to using 
$H_U + S_U$ where $S_U = -K\sum_{\Box}\prod_{\Box}U_{ij}$ instead of $H_U$. 
It is clear from the context that, in this case, the plaquettes under 
consideration belong to the primary lattice. If we now introduce the 
two-component vector $\Omega = (U, V)$ and the matrices 
\begin{equation}\label{MN}
  M = \left(\begin{matrix}
  0 & 1 \\
  1 & 0 
  \end{matrix}\right), 
\qquad 
  N = \left(\begin{matrix}
  K & 0 \\
  0 & \tilde{J} 
  \end{matrix}\right)
\end{equation}
the partition function can be rewritten in the neat form  
\begin{equation}\label{trace}
Z = {\hbox{Tr}}~ e^{-\beta (H_U + KS_U)} =  A(J) \sum_\Omega e^{-S} 
\end{equation}
where 
\begin{equation}\label{MN}
S = {\beta\over 2}\sum_{<ij>}i{\pi\over 4}(1 - \prod_\Box \Omega )M(1 - \Omega )
- \beta N\sum_{\Box}\prod_{\Box} \Omega_{ij}
\end{equation}
The $<ij>$ in the above equation refers to, as before, nearest neighbour 
sites, but these could now be either on the primary or dual lattice. Next we 
take the infra-red (low momentum) limit and drop the $N$-term to get,   
\begin{equation}\label{M}
S = {\beta\over 2}\sum_{<ij>}i{\pi\over 4}(1 - \prod_\Box \Omega )M(1 - \Omega )
\end{equation}
This is recognised as the lattice version of the topological Chern-Simons
theory encountered in the studies of the double layer quantum Hall 
effect \cite{zee}, and sometimes also referred to as the BF-theory \cite{bf}.

We are now in a position to examine the entanglement properties of Ising spins
in the above system. The object of central interest in studying entanglement 
is the (reduced) density matrix. Once it is known, a relatively straightforward
calculation yields the von Neumann entropy $S$ of the system through the 
standard formula $S = -{\hbox{Tr}}\rho{\hbox{ln}}\rho$. The single particle 
reduced density matrix $\rho_i$, obtained by tracing over all the spins except 
the $i$-th spin, can be expanded as $\rho_i = {1\over 2}\sum_{\alpha = 0}^3 
c_\alpha\sigma_i^\alpha$ where $\sigma^0 = {\bf 1}$ and 
$\sigma_i^{\alpha\neq 0}$ are the Pauli matrices at site $i$. The requirement 
that Tr$\rho_i = 1$, the reality of $H$, and the global spin-flip symmetry of 
the Hamiltonian imply that apart from $c_0$ which is unity, $c_1$ is the only 
non-vanishing expectation value. A general expression for the coefficients 
$c_\alpha$ reads 
\begin{equation}\label{corr1}
c_\alpha = {1\over Z}<\sigma_i^\alpha~~ {\cal P}\prod_{\Gamma(i, j = \infty)}
U_{ij}>  
\end{equation}
The correlation function has been modified by the path-ordered insertion of a 
string $\Gamma$ of links connecting the point $i$ to $\infty$ making it 
path-dependent; the modification ensures that the density matrix is 
gauge-invariant. In a similar fashion, we can obtain the two-particle reduced 
density matrix by expanding it in terms of the tensor product of Pauli matrices
at the two sites under consideration: $\rho_{ij} = {1\over 4}\sum_{\alpha , 
\beta = 0}^3 c_{\alpha\beta}\sigma_i^\alpha\otimes\sigma_j^\beta $.   
The coefficients of the expansion $c_{\alpha\beta}$ are the connected 
correlators of two Wilson lines. The two-particle reduced density matrix is 
useful in calculating the amount of entanglement {\it localisable} between the 
two chosen spins \cite{verst}. Similar results hold for three-particle (as in 
the GHZ-state discussed by Arvind) and, in general, for $n$-particle 
entanglement. In each of these cases, the density matrix is expressed in terms 
of correlators of gauge invariant observables in a topological field theory 
{\it i.e.} topological invariants \cite{note}. The difference between two 
choices of the path $\Gamma$ is a measure of the frustrations enclosed by the 
loop formed by the two paths. There is another way in which a closed loop can 
be obtained from an open path namely, by imposing periodic boundary conditions.
If we choose this option, $c_\alpha$ and $c_{\alpha\beta}$ are just the 
expectation values of Wilson loop observables in the lattice Chern-Simons 
theory. 

Although the Wilson loop observables are gauge-invariant, they are by no 
means the only interesting observables. Notice that the path connecting 
two points on the primary lattice pierces one plaquette on the dual lattice
with every step it advances, accumulating one unit of flux in the process.
Let us therefore consider the gauge-invariant operator $C = V^{-1}UV$ obtained
by dressing (conjugating) the string $U_{ij}$ by the group-valued fields $V$
on the dual lattice. Labelling the dual lattice sites by barred coordinates, 
if $V$ runs from site $\bar i$ to site $\bar j$, $V^{-1}$ runs in the opposite 
direction circumnavigating the link $U$. We can use this operator instead of 
the Wilson loop to define the reduced density matrices. It may be mentioned 
that because of duality, the above conjugation operation simply corresponds to 
local unitary transformations of nearest neighbour spins on the dual lattice. 
Physically the operator $C$ represents a tube of dual plaquettes whose axis 
lies on the primary lattice with fixed end-points.  

So far there is nothing quantum about the discussion of entanglement. Indeed,
the Ising spins we have considered take values $\pm 1$, much like classical
bits; they are not allowed to be in any superposition state. The density 
matrices that we considered are purely thermal in nature, obtained, as they 
are, from the Gibbs' measure. This kind of entanglement is called thermal 
entanglement \cite{bose}. By using the standard Suzuki-Trotter \cite{nishi} 
method, however, we can map the three-dimensional Ising spin glass to a 
two-dimensional Ising spin glass in a transverse magnetic field. The presence 
of the transverse magnetic field allows for transitions between the two 
classical states of the Ising spins and makes the system quantum mechanical. 
Care must be exercised in defining the spin-flip operation: $\sigma_z$ and 
$\sigma_x$ behave differently under gauge transformations because of their
noncommutativity \cite{nishimori}. Furthermore, the situation here is slightly 
more complicated because the Suzuki-Trotter mapping from a $d$-dimensional 
classical statistical mechanical system to the $(d-1)$-dimensional quantum 
system, requires the classical system to have different couplings along the 
missing (replica) dimension and the remaining dimensions. This makes the 
duality transformation technically a little more involved. The results in 
so far as the entanglement (now truly quantum in nature) are concerned follow 
the same pattern. The correlation functions that appear in this case are those 
of the quantum spins on a two-dimensional square sub-lattice of the original 
three-dimensional lattice whose third dimension acts as the discretised time 
direction. To summarise, the reduced density matrices that one is interested 
in, both for thermal and quantum entanglement, are expressed in terms of 
expectation values of gauge invariant operators in a topological gauge theory. 
They are therefore topological invariants.   
 
It is difficult to obtain a ready physical insight into the correlation 
functions that appear above. Let us therefore consider a simpler case. Recall 
that similar results hold for all abelian groups $Z_p$. In particular for 
$p\rightarrow\infty$, we have a $U(1)$ Chern-Simons gauge theory on the 
lattice and all the paths are continuous. In this case, it is well-known that 
the topological invariant produced by a computation of the correlation function
of a pair of Wilson loops is the Gauss's linking number. The operator $C$ we 
introduced, is the lattice generalisation of the operator introduced in the 
continuum BF theory by Cattaneo {\it et al.} \cite{cat}, and gives it a nice 
physical interpretation as a tube of dual plaquettes. Its correlation function 
gives the Alexander-Conway polynomial of the (possibly) knotted axis of a 
plaquette-tube, with fixed end-points on the primary lattice. The partition 
function $Z$ is also a topological invariant, namely, the Reidemeister torsion 
of the three manifold defined by the boundary conditions we choose to put on 
our lattice. It is equivalent to the corresponding invariant in the continuum 
theory, namely, the Ray-Singer analytic torsion \cite{adams}. Curiously enough, 
the entanglement properties of the GHZ-state which has a non-zero tripartite 
entanglement, but a zero bipartite entanglement cannot be accounted for, by the
simple linking number invariants. This property is exactly like the 
corresponding property of Borromean rings which, however, are known to be 
distinguished from a disjoint union of unlinked rings by a higher order 
topological invariant, namely the Massey triple product \cite{massey}. An 
abelian topological theory cannot produce this invariant. It is therefore 
necessary to treat the spins as genuinely nonabelian objects, like one is 
forced to in the presence of a transverse magnetic field. In view of these 
remarks, the intriguing parallels between the entanglement of quantum states 
and the entanglement of braids, knots and links, that first appeared in 
\cite{arvind, kauff}, are more realistic -- may be even natural and 
deep-rooted -- than hitherto expected. 

Intuitively, topological invariants are insensitive to the presence, or 
changes, in length scales; this being the only distinction between a lattice 
and the continuum, we could borrow the relevant topological invariants from 
the simple continuum $U(1)$ theory in the above discussion. Such a luxury is 
lost if the group under consideration is finite, {\it e.g.} $Z_2$ in the case 
of the Ising model. Unlike classical gauge symmetries which come from 
continuous local invariances, finite groups usually appear as remnants
of a continuous symmetry group which is spontaneously broken. In such 
instances, these groups are known to give rise to cocycles in quantum field 
theory. This is a reflection of nontrivial group cohomology. In the present 
case, the finite group appears because we are dealing with spin systems. 
The construction of topological theories with finite groups follows from 
a deep result due to Dijkgraaf and Witten \cite{dw}, who showed that the 
Chern-Simons actions for a finite group $H$ are in one to one correspondence 
with the elements of the cohomology group $H^4(BH, Z)$, $BH$ being the 
classifying space of $H$. The isomorphism $H^4(BH, Z)\approx H^3(H, U(1))$ 
further implies that these actions are algebraic 3-cocycles, $\omega\in 
H^3(H, U(1))$, which take values in $U(1)$. The presence of no-trivial 
group cohomology in general leads to non-trivial $G$ bundles and the 
path integral involves a summation over all possible bundles. Unlike in the 
$U(1)$ theory, in which a determinantal expression can be derived for the 
partition function \cite{schwarz}, the resulting invariants in the case of 
finite groups --  called the Dijkgraaf-Witten invariants -- are not expressible 
in terms of determinants. Analogues of linking numbers in topological gauge 
theories with finite gauge groups have also been worked out by Ferguson 
\cite{ferguson}. 

As the first closing remark, we mention that the details omitted in this letter
can be found in a longer publication under preparation. Second, we wish to 
point out that many interesting problems remain. The connections between 
quantum entanglement and more sophisticated topological invariants like the 
Jones polynomial, require a nonabelian generalization of the results of this 
letter. Similar investigations in two and four dimensions should produce 
interesting connections between quantum entanglement and other important
mathematical results in low dimensional topology like the intersection theory 
on the moduli space of Riemann surfaces, and Donaldson's invariants 
respectively. Finally, it is not an exaggeration to say that this letter 
offers a mere glimpse of a new vista which is beginning to unfold, on 
the relevance of finite group cohomology in the studies of entanglement in 
spin glasses. We hope to dilate on these issues, in the near future.  

\begin{acknowledgements}

I thank M. Panero and Siddhartha Sen for some incisive comments and also   
A. P. Balachandran, T. R. Govindarajan, I. Tsutsui, and A. Wipf for their
interest and discussions. It is a pleasure to acknowledge the hospitality of 
D. O'Connor and T. Dorlas of the Dublin Institute for Advanced Studies where 
this work was completed.  

\end{acknowledgements}


\begin{thebibliography}{}

\bibitem{lor} L. O'Raifeartaigh {\it The Dawning of Gauge Theory}, 
(Princeton University Press, 1997).

\bibitem{chuang} M. A. Nielsen and I. L. Chuang {\it Quantum Computation and 
Quantum Information}, (Cambridge University Press, 2001).

\bibitem{wilczek} F. Wilczek {\it Fractional Statistics and Anyon 
Superconductivity}, (World Scientific Publishing Company, 1990);
G. Marmo, B. S. Skagerstam, A. Stern and A. P. Balachandran {\it Classical 
Topology and Quantum States}, (World Scientific Publishing Company, 1991).

\bibitem{leinaas} J. M. Leinaas and J. Myrheim, Nuovo Cimento {\bf 37B}, 1 
(1977).  

\bibitem{jackiw} R. Jackiw {\it Topological Investigations of Quantized Gauge 
Theories}, (Les Houches Summer School on Theoretical Physics: Relativity, 
Groups, and Topology, 1983); R. Rajaraman {\it Solitons and Instantons},
(North-Holland, 1987). 

\bibitem{thouless} D. J. Thouless {\it Topological Quantum Numbers in 
Nonrelativistic Physics}, (World Scientific Publishing Company, 1998);
M. I. Monastyrsky {\it Topology in Condensed Matter}, (Springer Series in 
Solid-State Sciences, 2005).

\bibitem{arvind} P. K. Arvind {\it R. S. Cohen et al (eds), Potentiality, 
Entanglement and Passion-at-a-Distance, 53-59} (Kluwer Academic Publishers, 
1997).

\bibitem{ghz} D. M. Greenberger, M. A. Horne and A. Zeilinger {\it M. Kafatos
(ed), Bell's Theorem, Quantum Theory, and Conceptions of the Universe}
(Kluwer Academic Publishers, 1989). 

\bibitem{kauff} L. H. Kauffman and S. J. Lomonaco Jr., New Journal of Physics
{\bf 4} 73, (2002).  

\bibitem{witten} E. Witten, Commun. Math. Phys. {\bf 121} 351, (1989). 

\bibitem{neumann} J. von Neumann {\it Mathematical Foundations of Quantum 
Mechanics}, (Princeton University Press, 1996).

\bibitem{kaul} M. Atiyah {\it The Geometry and Physics of Knots}, (Cambridge
University Press, 1990); E. Guadagnini {\it The Link Invariants of Chern-Simons 
Field Theory}, (Walter de Gruyter, 1993); R. K. Kaul, T. R. Govindarajan, 
and P. Ramadevi, hep-th/0504100 {\it To appear in Encyclopaedia of Mathematical
Physics, Elsevier}.

\bibitem{dw} R. Dijkgraaf and E. Witten, Commun. Math. Phys. {\bf 129} 393, 
(1990). 

\bibitem{fradkin}{\it This seemingly innocuous generalisation has non-trivial 
repercussions. The competition between the two types of bonds leads to 
configurations which are frustrated. The frustrations are represented by 
elemental plaquettes for which the product of all the bonds is $-1$. They act 
like dislocations or points of curvature in an otherwise flat lattice and 
present topological obstructions to shrinking Wilson loops to  points, thus 
producing a $Z_2$-gauge Aharonov-Bohm effect. For details, see} E. Fradkin,
B. A. Huberman and S. H. Shenker, Phys. Rev. {\bf B18} 4789-4814, (1978). 

\bibitem{debu} D. Chowdhury {\it Spin Glasses and Other Frustrated Systems}, 
(World Scientific Publishing Company, 1986).

\bibitem{senthil} T. Senthil and M. P.A. Fisher, Phys. Rev. {\bf B63} 134510,
(2001).  

\bibitem{savit} R. Savit, Rev. Mod. Phys. {\bf 52} 453-487, (1980). 

\bibitem{adams} D. H. Adams, Phys. Rev. Lett. {\bf 78} 4155-4158, (1997).
 
\bibitem{deser} S. Deser, R. Jackiw, and S. Templeton, Phys. Rev. Lett. 
{\bf 48} 975-978, (1982); Annals of Physics {\bf 140} 372-411, (1982).

\bibitem{wegner} F. Wegner, J. Math. Phy. {\bf 12} 2259-2272, (1971).

\bibitem{sen} Samik Sen, Siddhartha Sen, J. C. Sexton, D. H. Adams, 
Phys. Rev. {\bf E61} 3174-3185, (2000).

\bibitem{kantor} R. Kantor and L. Susskind, Nuc. Phys. {\bf B366} 533-568, 
(1991).

\bibitem{kitaev} A. Yu Kitaev, Annals of Physics {\bf 303} 2-30, (2003);
M. H. Freedman, A. Kitaev, M. J. Larsen, and Z. Wang, Bull. Am. Math. Soc.
{\bf 40} No. 1, 31-38, (2002); J. Preskill, Caltech Ph219 Lecture Notes, 
Chapter 9, http://www.theory.caltech.edu/preskill/ph229 

\bibitem{delft} I. Chiorescu, Y. Nakamura, C. J. P. M. Harmans, J. E. Mooij,
Science {\bf 299} 1869, (2003); I. Chiorescu, P. Bertet, K. Semba, Y. Nakamura,
C.J.P.M. Harmans, J. E. Mooij, Nature {\bf 431} 159, (2004). 

\bibitem{nishi} H. Nishimori {\it Statistical Physics of Spin Glasses and 
Information Processing: An Introduction}, (Oxford University Press, 2001).

\bibitem{sourlas} N. Sourlas, Nature {\bf 339} 693, (1989).

\bibitem{zee} X. G. Wen and A. Zee, Phys. Rev. {\bf B46} 2290-2301, (1992).

\bibitem{bf} D. Birmingham, M. Blau, M. Rakowski, and G. Thompson, Phys. Rep. 
{\bf 209}, 129-340, (1991).

\bibitem{verst} F. Verstraete, M. Popp and J. I. Cirac, Phys. Rev. Lett.
{\bf 92} No.2 027901, (2004). 

\bibitem{note} {\it It is in general true that the reduced density matrices are
completely determined by the correlation functions of physical observables;  
these, however, are in general difficult to calculate exactly. For a 
representative list see,} T. J. Osborne and M. A. Nielsen, Phys. Rev.{\bf A66} 
032110, (2002); S. Bose, Phys.  Rev. Lett. {\bf 91} 207901, (2003); A. 
Lakshminarayan and V. Subrahmanyam, Phys. Rev.{\bf A67} 052304, (2003); 
V. Subrahmanyam, Phys. Rev. {\bf A69} 022311, (2004); Phys. Rev.{\bf A69} 
034304, (2004). 

\bibitem{bose} {\it In the field of critical phenomena, it has taken a 
curiously long time to coin the name {\it quantum phase transitions} for  
macroscopic changes of state driven by parameters other than temperature 
{\it e.g.} magnetic fields. In contrast, the realisation that entanglement 
between spins can be established by purely thermal - as opposed to quantum 
correlations - has dawned fairly early. See, for example} M. C. Arnesen, S. 
Bose and V. Vedral, Phys. Rev. Lett. {\bf 87} No.1 017901, (2001). 

\bibitem{nishimori} S. Morita, Y. Ozeki, H. Nishimori, J. Phys. Soc. of Japan
{\bf 75} No. 1 014001, (2006).

\bibitem{cat} A. S. Cattaneo, P. Cotta-Ramusino, M. Martellini, Nuc. Phys. {\bf
 B436}, 355-384, (1995).

\bibitem{massey} W. S. Massey, Journal of Knot Theory and Its 
Ramifications, {\bf 7} No. 3, 393-414, (1998).

\bibitem{schwarz} A. S. Schwarz, Lett. Math. Phys. {\bf 2} 247-252, (1978).

\bibitem{ferguson} Kenneth Ferguson, Journal of Knot Theory and Its 
Ramifications, {\bf 2} No. 1, 11-36, (1993).
\end{thebibliography}
\end{document}